\documentclass[conference]{IEEEtran}
\IEEEoverridecommandlockouts
\usepackage{cite}
\usepackage{amsmath,amssymb,amsfonts}
\usepackage{algorithmic}
\usepackage{graphicx}
\usepackage{textcomp}
\usepackage{xcolor}
\usepackage{float}
\def\BibTeX{{\rm B\kern-.05em{\sc i\kern-.025em b}\kern-.08em
    T\kern-.1667em\lower.7ex\hbox{E}\kern-.125emX}}
\begin{document}

\title{Implementing A Middleware API for Facilitating Heterogeneous IoT Device Communication Protocols and Data Retrieval}

\author{
\IEEEauthorblockN{Sasoun Krikorian}
\IEEEauthorblockA{\textit{Computer Science} \\
\textit{Worcester Polytechnic Institute}\\
Worcester, United States \\
smkrikorian@wpi.edu}
\and
\IEEEauthorblockN{Benjamin Skarnes}
\IEEEauthorblockA{\textit{Computer Science} \\
\textit{Worcester Polytechnic Institute}\\
Worcester, United States \\
bcskarnes@wpi.edu}
\and
\IEEEauthorblockN{Sai Varun Vadlamudi}
\IEEEauthorblockA{\textit{Computer Science} \\
\textit{Worcester Polytechnic Institute}\\
Worcester, United States \\
svadlamudi2@wpi.edu}
}

\maketitle

\begin{abstract}
Currently, there are over 14 billion IoT devices \cite{num iot devices}, and with many devices come many protocols, the main ones being MQTT and CoAP. We are interested in connecting the many diverse IoT devices to the cloud. To do so, we use the middleware architecture proposed by article \cite{garg and dave}  in which a device, called the middleware, acts as the middleman between the various IoT networks and the cloud. 

Since IoT devices typically operate in real-time, performance is of great concern. Therefore, we conducted a simulation to measure the data latency of using middleware and the overall fairness between different IoT networks. Our simulation had an MQTT and a CoAP network interacting with the middleware. The simulation results showed that CoAP always had a lower travel time than MQTT, mainly because CoAP is a more lightweight protocol. However, we also found that MQTT had slightly more throughput, which was unexpected since we initially thought that CoAP would have had higher throughput. 

We have shown that analyzing data via a middleware device is possible and that there are potential directions to explore, such as evaluating different Quality of Service Algorithms in the context of having a middleware device. 
\end{abstract}

\section{Introduction}
\subsection{Background}
The Internet of Things, or IoT, is an emerging paradigm that aims to bring humanity into the next generation of a high-tech lifestyle. IoT aims to connect smart devices to the internet to provide efficiency, connectivity, and an overall higher quality of life for its users. These devices can range from intelligent wearables, fitness devices, and smart utilities, such as thermostats, to even entire smart homes. 

A few of the most widely used communication protocols include MQTT and CoAP, which are specific to IoT devices that match their requirements as resource-constrained devices. MQTT and CoAP are widespread because they scale and are modular \cite{mqtt vs coap}. Scalability in the context of IoT is essential because an IoT network can contain many IoT devices, and having the devices work in real-time is a crucial feature.

Message Queue Telemetry Transport, or MQTT, is a lightweight and scalable publish/subscribe IoT messaging protocol that can be easily implemented to relay data through a central server, also known as a broker. In the context of MQTT, the broker acts most like a post office. The client (IoT device) sends a message to the broker (server), and then the message is forwarded or “published” to everyone who is ``subscribed'' to this particular topic of message \cite{b3}.

The Constrained Application Protocol, or CoAP, is an application-layer protocol developed as a highly lightweight communications protocol for resource-constrained devices. CoAP uses the GET, PUT, POST, and DELETE methods and response codes similarly but not precisely like HTTP. CoAP uses the resource model mapped to the Universal Resource Identifier instead of the topic-based system that MQTT operates on \cite{b3}.

MQTT generally uses a publish/subscribe-based model, whereas CoAP uses a request/response-based model. MQTT can support multiple publishers and subscribers, whereas CoAP is a one-to-one protocol. The other main difference is that MQTT necessitates a reliable, lossless, in-order transport that typically requires the TCP/IP stack. On the other hand, CoAP runs on top of User Datagram Protocol, or UDP \cite{b3}.

This is where middleware comes into play. In 2018, the market for IoT middleware was valued at 6.44 billion dollars and is expected to reach 18.68 billion dollars by 2024. Middleware can be described as a software layer between the applications and the ``things'' of the Internet of Things. It is used to hide the heterogeneity of IoT devices and provides solutions to problems mentioned above, such as interoperability \cite{cruz rodrigues}\cite{behara}.

Roy Fielding introduced REST in his dissertation. REST stands for REpresentational State Transfer. REST is guided by a set of constraints that an API must follow \cite{REST}. The advantage of REST API is that the server becomes more accessible to manage from the constraints. Enforcing the stateless constraint is helpful since the user will be guaranteed to receive the same response if they send the same data. If the server were not stateless, the user could get a different reaction, making debugging more challenging, as reproducing errors could now take many more steps. Also, statelessness helps with ensuring that the code is data-driven. Also, the statelessness constraint can allow the user to cache the server's response since they know that sending the same data will result in the same response. With the cache constraint, the user can know whether or not the response can be cacheable, and knowing when to cache is essential for optimization purposes. The uniform interface constraint simplifies the API on both the provider and the user end since, rather than figuring out which API sends a GET request and which one sends a POST request, the same API can handle different requests. 

\subsection{Problem Description}
The IoT device network is expanding far and wide and is projected to have 125 billion devices connected over the next ten years \cite{b1}. With the explosive growth of IoT, many kinds of devices exist for many types of use cases. For example, one might want a door sensor to trigger a smart alarm to sound when an intruder enters. However, the door sensor and the smart alarm may come from different brands, so they may not know how to communicate with each other. Since there are numerous IoT devices, they are frequently heterogeneous; often, they cannot communicate with each other as they speak different ``languages''. 

Furthermore, there is a lack of standardization for how IoT devices communicate, which decreases the interoperability of devices and the ability to gather and analyze data on large numbers of heterogeneous IoT devices \cite{b4}. A few platforms out there work to translate the different protocols and provide analysis, such as Hadoop, Azure, and SmartThings; however, they are closed ecosystems, and any automation needs to be done through their platform.

\subsection{Limitations of Current Approaches}
Article \cite{garg and dave} proposes a middleware architecture to handle heterogeneous applications that will ``become the bridge between things and applications in the cloud''. Their middleware API is responsible for IoT device registration. Specifically, Figure 5 in the journal describes the following process. A user authorizes an IoT device by registering it through an exposed middleware API. The IoT device is sent the access token once registration is complete. Upon receiving the access token from the middleware API, the IoT device can retrieve data from a source via the access token. This proposed solution provides a middleware architecture that takes the responsibility for device registration, identification, and database management of the IoT devices. In this architecture, data sent from the IoT device is encrypted end-to-end since the communication between the gateway and the middleware API is encrypted through traditional practices. To expand upon this research, we seek to use the exposed middleware API, not for authentication but for further interaction with heterogeneous IoT devices.

\subsection{Challenges}
IoT devices come in various forms, with different communication protocols and data formats. For example, \cite{raschbichler} shows that MQTT treats the payload as a raw sequence of bytes and can be any arbitrary data, such as strings, JSON, or binary data. As a result, managing the diversity of IoT communication protocols and creating a standardization for how the data should be inserted into the backend database can be difficult. We handled this by parsing the data from the IoT devices inside our gateway and formatting it into standardized JSON for use in our REST API.
In addition, IoT applications often require real-time or near-real-time data processing, and meeting the latency requirements with a middleware implementation can be difficult.\cite{amoros} One way to improve latency is by selecting an adequate Quality of Service algorithm.
Finally, as the number of IoT devices increases and the traffic coming into the middleware increases, it may be difficult to implement scalability measures to ensure that data is correctly sent to the database without compromising performance. One possible solution could be implementing a cloud computing model in the middleware where resources are dynamically scaled to fit the processing requirements.\cite{iotsizematters}

\subsection{Summary}
In our simulation, we had 5 MQTT devices and 5 CoAP devices, where each device sent a message to the gateways at 1-second intervals, and the middleware API was invoked. Approximately ten messages were being sent every second, five being over MQTT, the other 5 being over CoAP. We had each message come with a timestamp to track the data latency and throughput of the packets. 

In Figures 5 through 7, we plotted the travel times of the messages sent using MQTT and CoAP over different periods. As each message was being sent from a device in MQTT, the travel time increased for each message. Due to the increased travel times, we see a sawtooth pattern in the plot. In contrast, the travel time remains relatively constant when looking at messages sent via the CoAP protocol. These results are expected, since CoAP is designed to be more lightweight than MQTT. 

We also plotted the distribution of messages received by the database for both communication protocols over various periods. We expected the distribution of messages transmitted to be split equally; however, we saw that the number of MQTT messages received by the database was slightly higher than the number of CoAP messages received. These results contradict our initial assumption that CoAP should have a higher throughput because it is a more lightweight protocol than MQTT.

Overall, we have achieved our goals of researching the latency and throughput of IoT networks utilizing a middleware device to create a platform where heterogeneous IoT devices can communicate. Furthermore, we also showed that data analysis through a middleware API is possible. Regarding communication, with our framework and gateways accessing the middleware, a CoAP client can quickly send a GET request to retrieve data inserted by an MQTT publisher and vice versa, effectively allowing communication between heterogeneous devices.

As a future work, one could investigate the different variations of network distributions by sending various amounts of messages from MQTT and CoAP. Another potential area of research is testing different Quality of Service (QoS) algorithms. The QoS algorithm we used is ``Fire and Forget’’, but other algorithms have different performance and reliability guarantees, and seeing how the distribution of messages changes based on the QoS algorithm being used would be interesting.

\section{Literature Review}
\subsection{Databases: SQL vs NoSQL}
SQL and NoSQL databases are the two central databases that handle IoT data. SQL databases use the relational model to store data, using tables with rows and columns and linking different tables to show relations between them. In contrast, NoSQL databases store data in files as documents or graphs. Since SQL databases hold data in separate tables, combining data requires JOIN statements, which are time and performance-intensive. In NoSQL, data is stored as objects that contain all related data, removing the need to combine data from different tables\cite{SQL}.

SQL and NoSQL database implementations prioritize different qualities in their designs. SQL prioritizes consistency, while NoSQL prioritizes availability. They also differ in how they scale. SQL databases support vertical scalability, meaning that you can improve the performance by enhancing the processing power of the machine that the database is on. NoSQL supports horizontal scalability, meaning you can spread the database over multiple machines, allowing the high availability principle\cite{SQL}.

When testing the performance of different database types, research shows little difference in performance when dealing with IoT data; however, this research does not use a timescale database implementation, which could affect the different architectures differently \cite{SQL performance}.

\subsection{IoT Middleware}
The topic of IoT middleware is not new, and there have been multiple research areas on the topic of using middleware to assist in various functions, such as increasing security, scalability, or, in our case, interoperability.

Garg and Dave describe IoT middleware as an additional layer between IoT devices and the cloud applications that reduces computation on the cloud \cite{middlewaresec}. In a typical IoT device model, the middleware connects to different devices and handles computation or data manipulation before the information is pushed to the cloud. It is important to note that the data stored in the middleware is sensitive and private to the IoT device sending it. Therefore, the middleware must have security measures to prevent unauthorized users from accessing confidential information. The article introduces Attribute-based encryption (ABE) on the middleware for access control. According to the research, this proposed scheme aims to provide security and efficiency while reducing the complexity of middleware. This is one example of how middleware is being used amongst IoT devices. Our research seeks to move beyond security and focus on interoperability between IoT devices with diverse communication protocols. We are not so concerned with the security of the information being passed from one IoT device to the next but with the ability to analyze heterogeneous IoT devices through a middleware platform.

In addition, middleware is also being used to tackle the issue of scalability in the IoT space. \cite{middlewareclust} states that single middleware will suffer from limited computing and memory capacity when the number of publishers and subscribers increases. Therefore, the article proposes creating a middleware cluster that provides services to more transactions. Their preliminary test results showed that the number of concurrent messages per second was up to 30 for CoAP and 8 for MQTT. This again shows the usage of middleware to tackle some of IoT’s most pressing issues, which, in this case, was scalability. This research only examines the performance of middleware where the data is being pushed to a web client for analysis. However, we want to expand upon this to get information to treat the middleware as an API to allow various external applications or users to analyze the data from heterogeneous IoT devices on a network.

\begin{figure}
   \centering
   \includegraphics[width=0.5\textwidth]{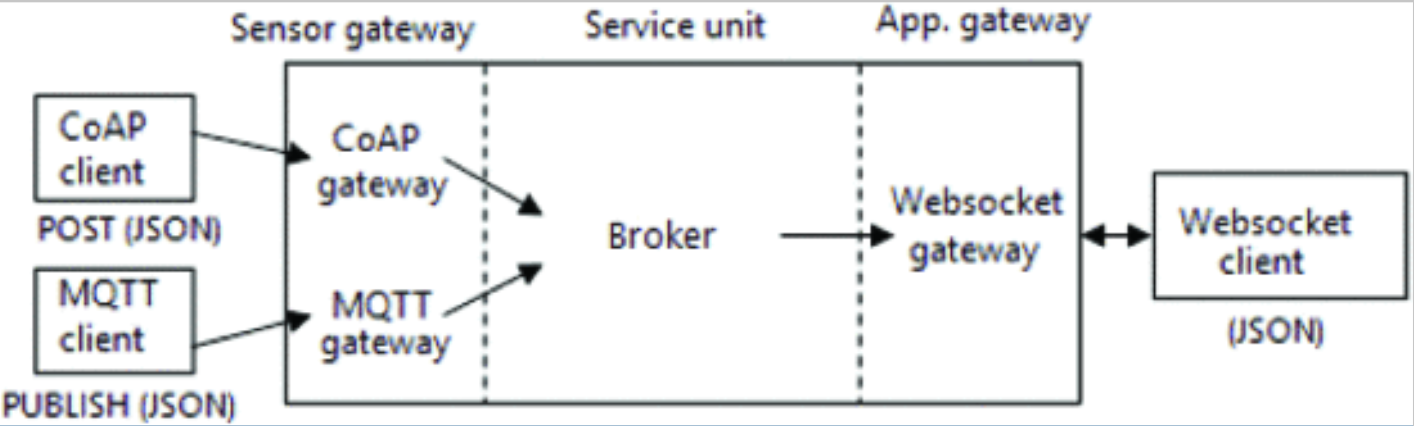}
   \caption{Middleware Architecture for Interoperability}
   \label{fig:middlewareinter}
\end{figure}

Finally, middleware is also being used to address the very problem this research aims to address: interoperability. \cite{middlewareinter} states that along with the increasing number and diversity of devices connected, there arises a problem of interoperability. Specifically, the article talks about syntactical interoperability, where the IoT should be able to connect all devices through various data protocols. The article proposes a middleware capable of acting as a gateway between CoAP, MQTT, and WebSocket. The proposed architecture is implemented using the “Ends to middle” model and can be seen in figure \ref{fig:middlewareinter}. Messages coming in go to separate gateways based on whether they are using MQTT or CoAP, and then this information is processed in the central service unit or the middleware. The middleware then sends this information to the web client. While this is very similar to what we aim to achieve, this research does not offer an API that can be used for data analysis on heterogeneous IoT devices.

\section{Methodology}
In the following section, we discuss certain technologies used within this research and answer why and how we are using said technologies. Specifically, we focus on the three central portions of the architecture: simulation of IoT Devices, API Development, and Databases.

\subsection{Simulation of IoT Devices}
The complexity and volume of IoT devices on a shared network are rising. As a result, this could exponentially increase the chance of an unanticipated problem that could collapse the network as a whole. The flexibility and scalability that make IoT deployments attractive also make them difficult to test. However, this is where simulation plays a crucial role. Simulation is essential when the number of connected devices grows beyond a handful. When there are so few, it makes sense to configure them manually, but when you have hundreds, it quickly becomes infeasible. Simulation is a far more efficient way of handling and testing large numbers of IoT devices. Still, simulating an extensive network is also far more cost-effective than building one. For these reasons, we thought it best to employ a simulated network rather than create one of our own with physical IoT devices because we seek to build an extensive network with many devices, and through simulation, it will be far more efficient and cost-effective \cite{iotsimulation}

\subsection{Network Simulators}
There are currently many network simulators on the market, but the main problem with many is that they are too complex and have far too advanced built-in tools for our simple research. We aim to create IoT devices that communicate over MQTT and CoAP and send these messages to a shared server that handles these requests. After coming across services like AWS IoT Simulator, IoTify, SimpleSafe, and many others, they all had the same underlying problem described above. To delve more into this, let’s look at just one: SimpleSafe. SimpleSafe provides its SimpleIoTSimulator, an easy-to-use IoT sensor/device simulator that quickly creates test environments of thousands of these IoT devices on one computer. In addition, SimpleIoTSimulator supports a wide range of standard IoT communication protocols, including MQTT and CoAP, which we are particularly interested in for this research project. SimpleIoTSimulator allows thousands of IoT devices to run, test, and visualize on its GUI interface. However, this is not a good fit for our research because it already provides a default broker to publish the messages. In addition, there is little documentation describing how we can receive messages from a broker and perform complex functions. Within our research, we are trying to create a middleware API that can act as a broker for IoT devices that use MQTT and as a server in the client-server model of IoT devices that use CoAP. We aim to create a custom client connection to an MQTT broker and perform as a server for CoAP requests/responses. SimpleIoTSimulator from SimpliSafe and similar tools have a world-class simulator engine designed for scalability and functional testing focusing on individual communication protocols. In contrast, we are researching interoperability between multiple communication protocols through a universal middleware platform \cite{simplesoft}.

\subsection{Simulating MQTT}
Due to existing network simulators, we came across being too complicated to configure a common custom client-to-broker for MQTT and client-to-server for CoAP; we felt it was better to use existing Python libraries to simulate a simple IoT device and a middleware API that acted as the broker/server that precisely fit our needs.
We needed two main components, the broker and the client, to configure our own MQTT network without the aid of existing services like AWS IoT simulator, IoTify, or SimpleIoTSimuluator.

After researching a few MQTT brokers, we found the HiveMQ and Mosquitto message brokers. HiveMQ provides a free public HIveMQ MQTT broker that anyone can use. To access it, you give the broker address of ‘broker.hivemq.com’ and specific port information. On the other hand, Mosquitto MQTT is a downloadable library, and an MQTT broker can be hosted on your local machine simply by running the ‘moquitto’ command on a terminal. For our research, we chose to go with Mosquitto because we felt it best to keep everything local so that it is consistent and we are aware of all the processes being run \cite{hivemq} \cite{mosquitto}.

Through extensive research, paho-mqtt is arguably the best Python MQTT open-source client library. It was developed under the Eclipse Foundation. Paho-mqtt provides a detailed explanation for quick implementation and has an extreme community sport that will be vital to overcoming hurdles within our research. In addition, implementing simple MQTT clients typically requires only a few lines of code, which is also crucial under the strict time limitations of this research \cite{you}.

\subsection{Simulating CoAP}
For the abovementioned reasons, we also used Python libraries to simulate IoT devices that operate over the CoAP protocol. For CoAP simulation, we needed to simulate two main components: the client and the server. 

In choosing a Python library capable of simulating network devices that communicate over CoAP, the best one that precisely aligned with our research was aiocoap. The aiocoap package implements CoAP and is written in Python 3. Aiocoap can set up a server and a client and allow them to communicate over CoAP. And has a strong foundation and an abundance of documentation. Therefore, due to the ease of implementation and the strong community, we choose aiocoap as the Python library for simulating network devices over CoAP \cite{aiocoap}.

\subsection{Database Implementation}
Since we would like to support various IoT devices with many different types of data, the simplest way to store it is in time series string format. This will simplify the schema and store the data in one large hypertable. Since each IoT sensor device usually only sends one value over the protocols, we don't need to parse the data and can store it as-is.

\begin{figure}
   \centering
   \includegraphics[width=0.5\textwidth]{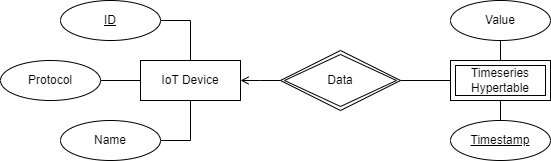}
   \caption{Database ERD}
   \label{fig:ERD}
\end{figure}

Since we don't need NoSQL databases' high availability and horizontal scalability, we will use TimeScaleDB. This SQL-type database is an extension that runs on top of PostgreSQL. We chose TimescaleDB because it is open-source and uses the same syntax as PostgreSQL, with which we are familiar. 

Figure \ref{fig:ERD} shows an example Entity Relation Diagram for the implementation, with one table holding the data from each device and assigning it as an ID, and another table, which will be created as a time series hypertable, will store the actual values from the IoT devices and have an additional ID field which references the devices.

\subsection{Middleware API Implementation}
We used a REST API for users to interact with the middleware to add devices and retrieve data. 

As mentioned, to implement our REST API, we used FastAPI. FastAPI is a Python library used for building API endpoints. With FastAPI, we could ensure that the API we created adheres to the constraints of REST. 

To ensure that our API is stateless, our responsibility will be to ensure that when the server needs information, it is only from the payload and nothing else. While statelessness makes resource management more manageable, the server cannot provide sessions, allowing the user to access sensitive data. However, the workaround to the issue would be to use tokens. The client can be issued a token, and the server can verify the token is valid. One can also set an expiration time on the tokens, which could violate stateless constraints since the server would now be tracking how old the key is. Having a token expiration is essential for security, as the attacker could potentially steal the token and use it to get access to sensitive information. To implement token expiration without violating REST, the workaround is JWT. A JWT is a signature of the JSON object, so any changes made to the payload will cause a change in the signature, meaning we will be able to. Fortunately for us, FastAPI supports using OAuth2 with JWT bearer tokens and sufficient documentation, which means we should be able to set up tokens right out of the box \cite{JWT}. GET requests, by nature, are cacheable. Other requests cannot be cached. The cache constraint is easily met. Uniform interface is our responsibility. For a given functionality, we must ensure that  GET, POST, etc., use the same endpoint for the provided functionality. Code on demand is not essential since IoT devices will exchange information, not render dynamic webpages. Since this constraint is optional, we can disregard it yet still be able to call our API a RESTful one. 

OpenAPI is a standardization of how to describe an API in a JSON or YAML file \cite{OpenAPI}. One creates an API description file that adheres to the OpenAPI standard. That way, the API is agnostic to any language, and anyone who wants to use the API will be able to understand the API without knowing the details. Also, since the description adheres to the OpenAPI standard, it will be machine-readable. Hence, the API becomes easy to mock and test, which could be helpful when simulating our IoT network. OpenAPI is beneficial for our purposes since it allows new compatible gateways to understand how the API works, rather than having to tediously configure each IoT gateway to be able to make an API call, while also giving us the flexibility to make modifications to the API without having to reconfigure all the IoT gateways again manually. With just ten or more IoT devices, constantly reconfiguring is already an error-prone and tedious task. Instead, the middleware can send a copy of the API description file, and the oncoming IoT can adjust accordingly, thus reducing unnecessary manual labor. 

One significant advantage of FastAPI over Flask or Django is that it is based on the OpenAPI specification. At first, we initially thought we would have to painstakingly write the API description file for each new API we create, but fortunately, there is a much better and faster way. Since FastAPI is based on OpenAPI, we can generate our API description file when creating a new API file. That way, we will not have to manually write our API description files, and we can delegate the task to FastAPI. 

\subsection{Experiment}
\subsubsection{Flow}
To conduct this experiment, IoT devices were simulated using the Python libraries described above. Some IoT devices communicated over CoAP, and some communicated over MQTT. In addition to this, we also had two gateways that acted as channels for the MQTT and CoAP messages. These gateways fed into a service unit or the middleware using API calls. The middleware then pushed messages to a database table consisting of the following attributes: mode message. The mode will be whether the message came in over MQTT or CoAP communication protocol, and the message will be the contents of the actual message. For our research, we fill the contents with the time the IoT device sends the message to the middleware. The middleware will also act as a REST API for users and applications to interact and analyze the heterogeneous IoT devices. Through this experiment, we aim to research the following questions:
\begin{itemize}
    \item What is the data latency when dealing with a middleware API accessed by IoT devices?
    \item How does the distribution of a heterogeneous network affect the throughput on a middleware API?
\end{itemize}

\begin{figure}
   \centering
   \includegraphics[width=0.5\textwidth]{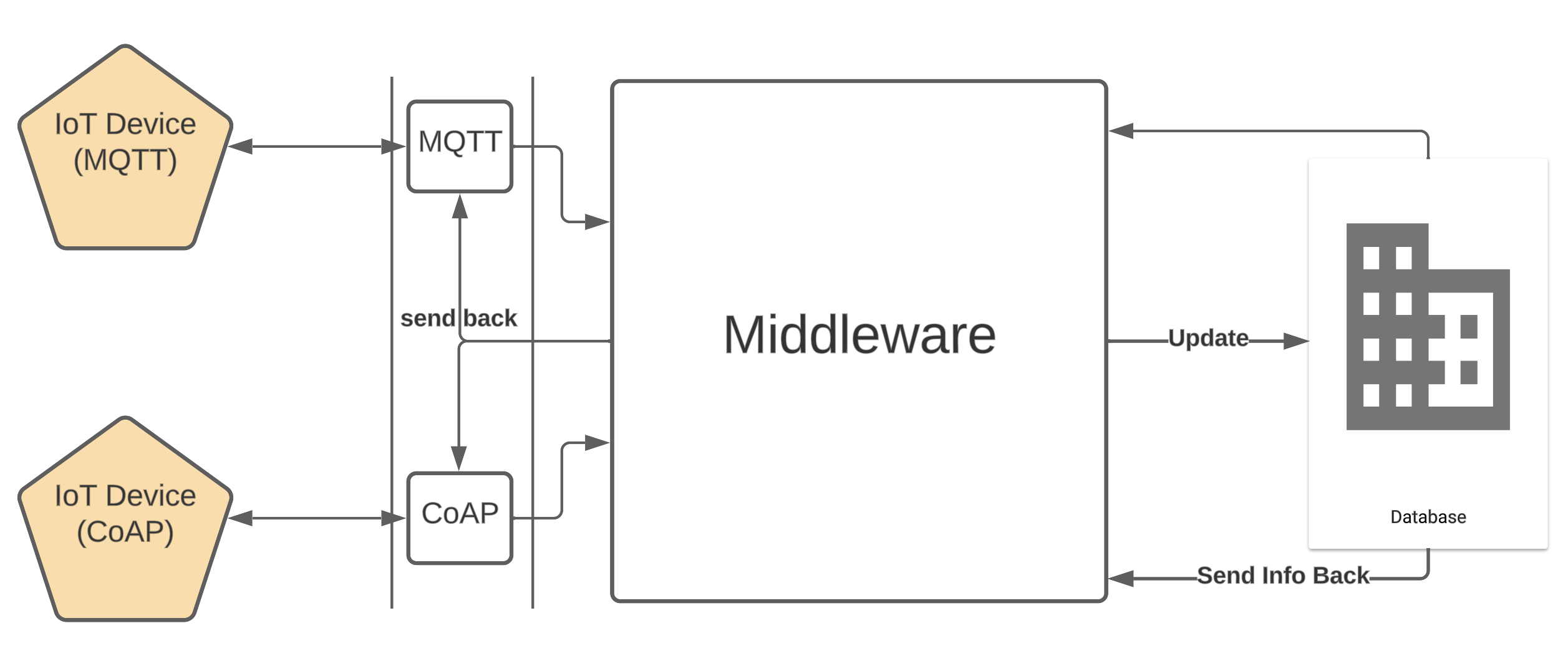}
   \caption{Overall Architecture}
   \label{fig:ourmiddlewarearchitecture}
\end{figure}

\subsubsection{Implementation}

The first thing we did was set up the IoT devices and have them communicate messages to the middleware. To do this, we set up MQTT and CoAP devices. We implemented a broker, publishers, and subscribers to set up an MQTT device. As mentioned above, we used the mosquito broker and the paho-mqtt library to simulate publishers and subscribers. One MQTT publisher acted as the device, and one MQTT subscriber was the gateway that forwarded the messages to the middleware.

As mentioned above, the CoAP devices utilized aiocoap to simulate a client and a server that sent messages to each other. In this case, the client is an IoT device that sends messages to the server that acts as the gateway, which then forwards the messages to the middleware.

We set up a database locally on a personal machine with the TimescaleDB extension for PostgreSQL and simulated the same machine that we ran the database.

Our network could be simulated on the cloud or locally; we chose to run our simulation locally as managing resources was significantly more straightforward for us. With FastAPI, we ran a uvicorn server locally, which is responsible for handling API calls. In other words, the uvicorn server was our virtual middleware. Then, we could have an external application to make API calls to the middleware. That way, we can evaluate network performance under stress. 

We evaluated the latency of the middleware by measuring the time it takes for an IoT device to send its data to the database. We accomplished this by having the IoT device place a timestamp on the packet. That way, when the database received the packet, it generated its timestamp and compared the current timestamp to the previous one. We could then determine the travel time of the packet from the IoT device to the database through the middleware API. 

We evaluated the fairness of the middleware by observing a time section of the database for 5, 10, and 30 seconds. From each entry, we determined whether the entry originated from the MQTT network or the CoAP network. Ideally, we wanted the database entries to be split fairly between the heterogeneous network and the size of each network. 

\section{Results}
In this section, we go over the empirical results and any conclusions that can be drawn from them.
\begin{figure}[ht]
   \centering
   \includegraphics[width=0.5\textwidth]{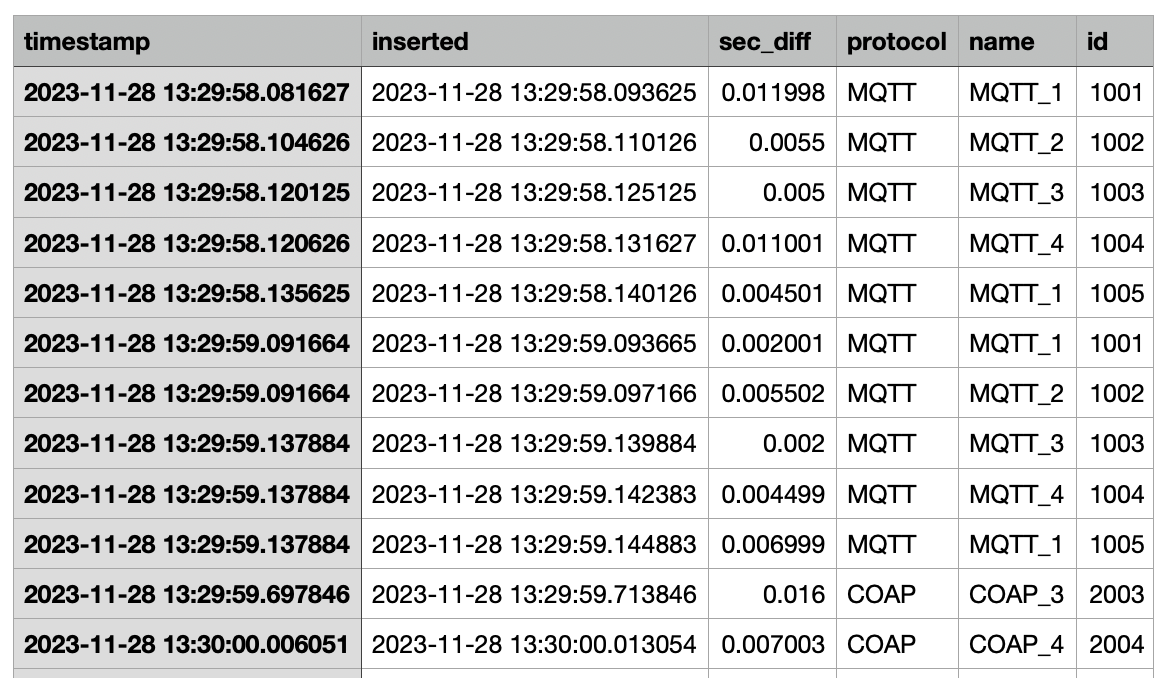}
   \caption{Example of Data}
   \label{fig:dataexample}
\end{figure}
Figure \ref{fig:dataexample} shows an example snippet of the data collected by the IoT device simulators. We had a script that handled 5 MQTT devices and another that took 5 CoAP devices. The scripts were run simultaneously, and the five devices asynchronously sent a message at 1-second intervals to the gateways, where the middleware API was then called. Therefore, the IoT devices sent ten messages per second; five were over MQTT, and 5 were over CoAP. Our middleware API had a POST function to insert the passed data into the database. The ``timestamp'' indicates the time that the message was created by the IoT device, and ``inserted'' represents the time the database received that message. That brings us to ``sec\_diff'', which is the difference in time between when the IoT device created the message and when it was acquired by the database, and in other words, it is the travel time of the IoT message through the middleware. In addition to this, we also noted the protocol, MQTT or CoAP, that the message was sent over. 
\begin{figure}[ht] 
   \centering
   \includegraphics[width=0.5\textwidth]{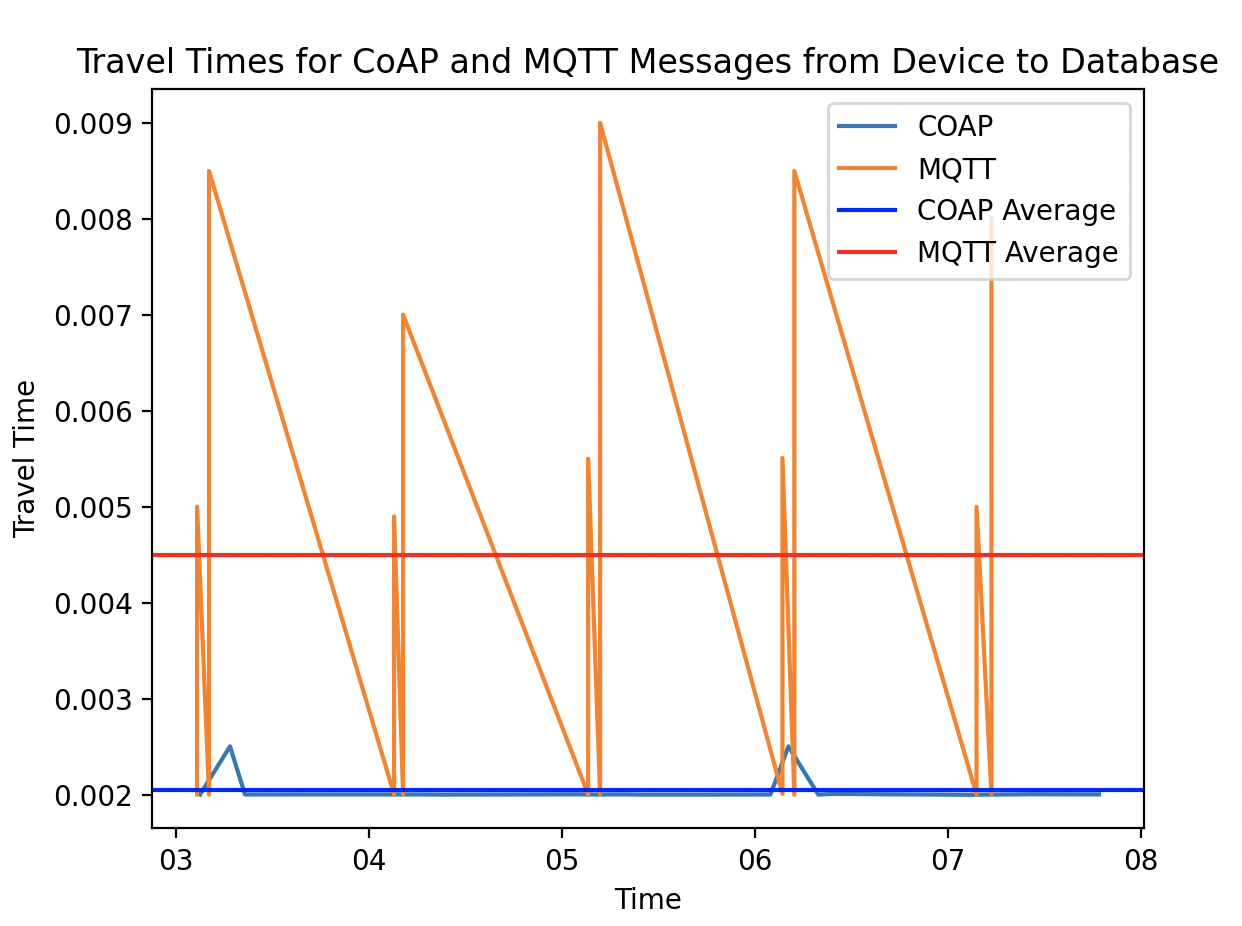}
   \caption{Travel Time of Protocols for 5 Seconds}
   \label{fig:traveltime5s}
\end{figure}
\begin{figure} [ht]
   \centering
   \includegraphics[width=0.5\textwidth]{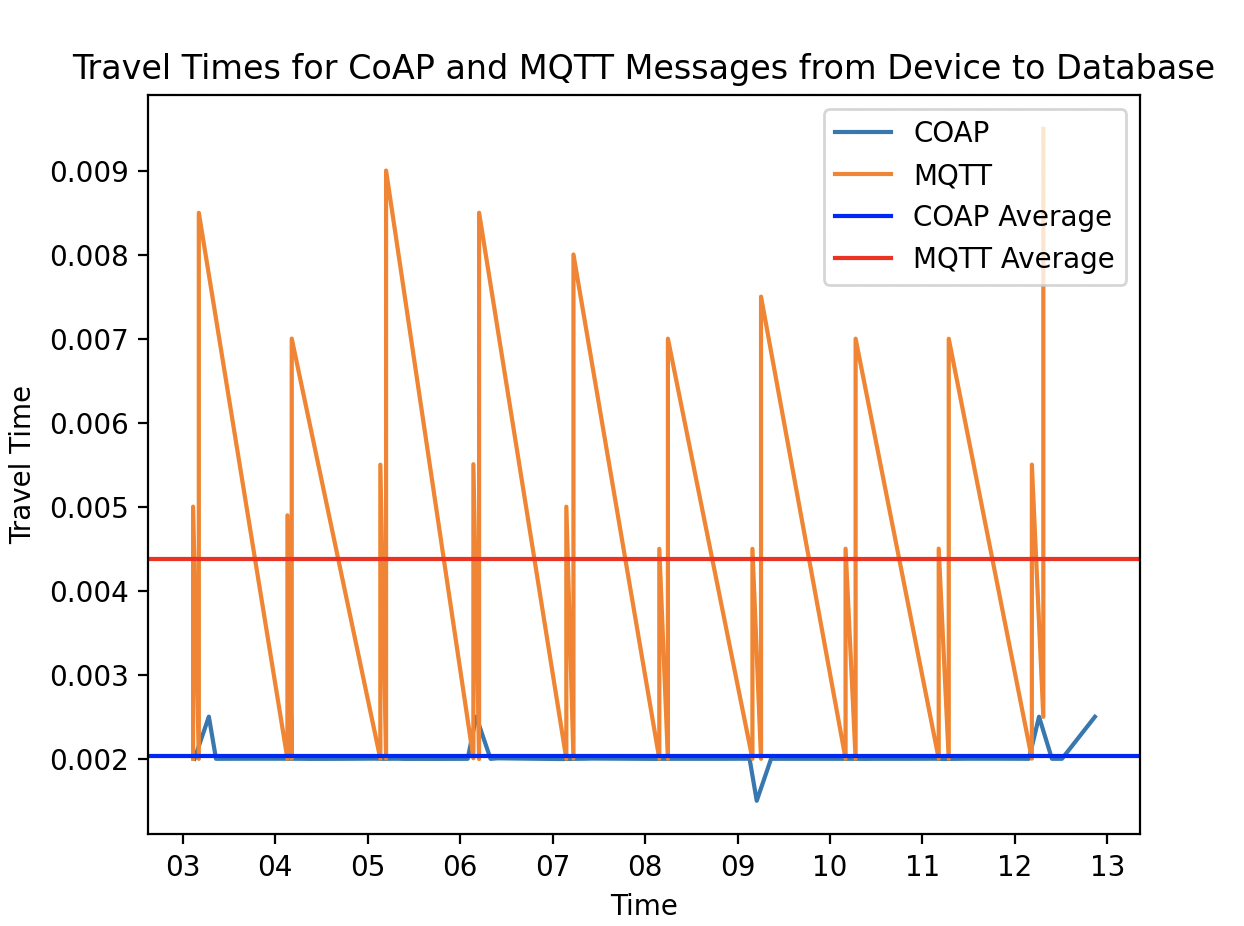}
   \caption{Travel Time of Protocols for 10 Seconds}
   \label{fig:traveltime10s}
\end{figure}
\begin{figure}[ht] 
   \centering
   \includegraphics[width=0.5\textwidth]{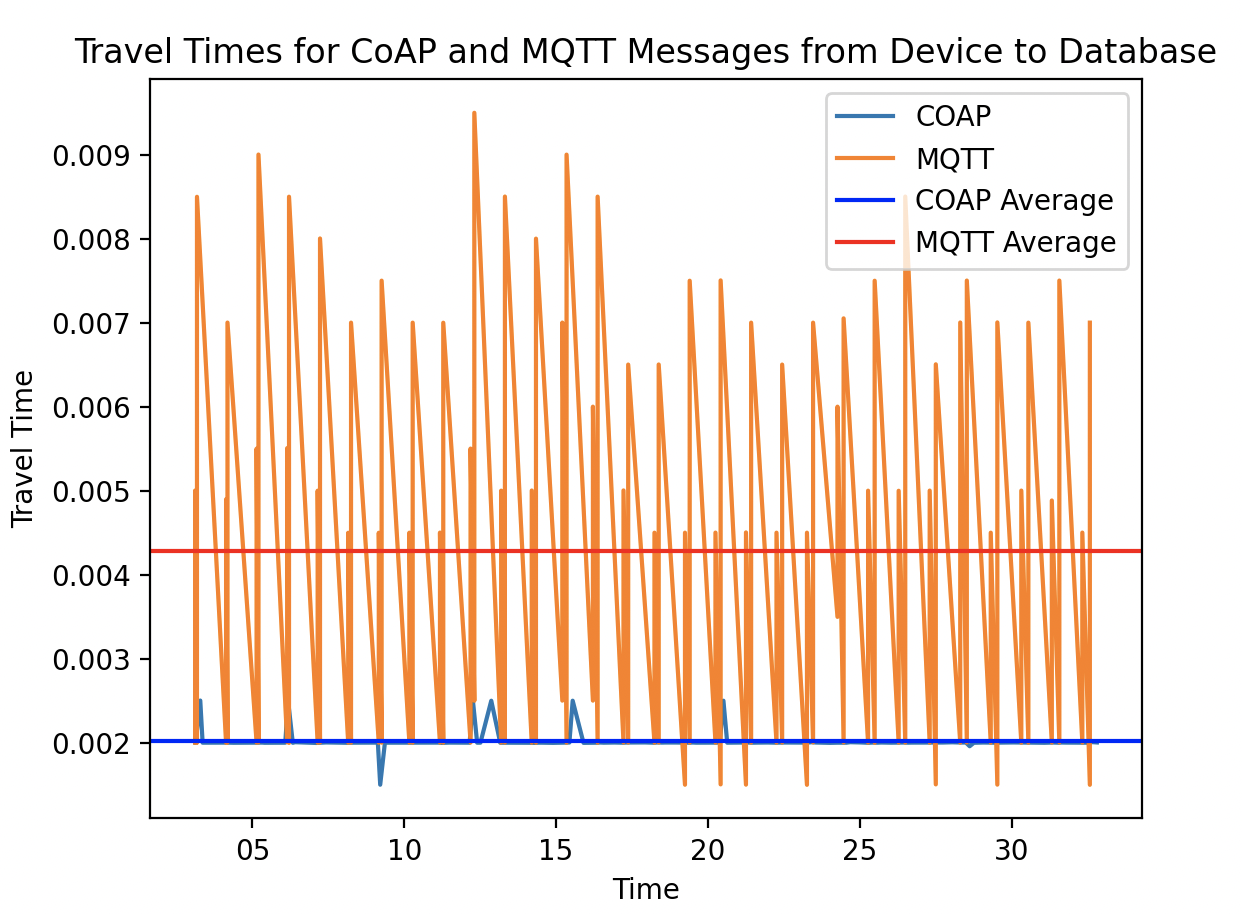}
   \caption{Travel Time of Protocols for 30 Seconds}
   \label{fig:traveltime30s}
\end{figure}

Figures \ref{fig:traveltime5s} through \ref{fig:traveltime30s} show the travel times of messages sent over MQTT and CoAP protocols over different periods they were recorded. Immediately, it was clear that when using the MQTT protocol, as messages were being sent, the travel time increased for each message. Then, when the messages were finished transmitting, the travel time decreased until the next wave of messages were transmitted. This behavior is what produced the sawtooth-like pattern. However, when looking at the messages sent through the CoAP protocol, we can see the travel time remains relatively constant at about 0.002 seconds. These results align with our initial assumptions of CoAP being a more lightweight protocol for constrained devices and, as a result, having a lower travel time through the gateway and middleware than MQTT.

\begin{figure}[ht] 
   \centering
   \includegraphics[width=0.5\textwidth]{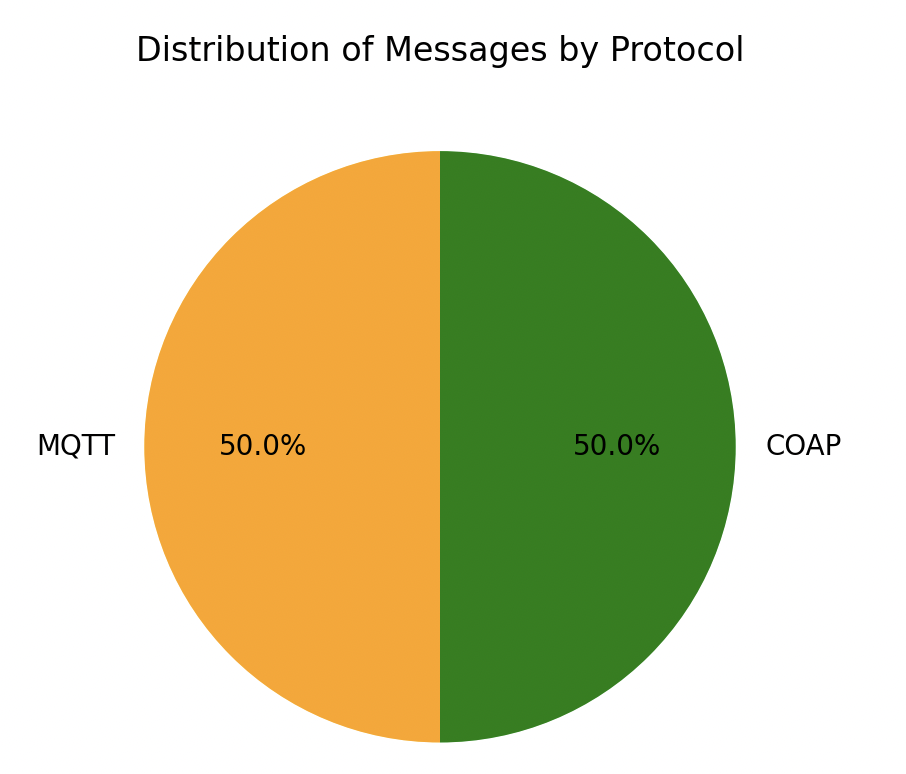}
   \caption{Distribution of Received Messages, 1 Second}
   \label{fig:distribution1s}
\end{figure}
\begin{figure}[ht] 
   \centering
   \includegraphics[width=0.5\textwidth]{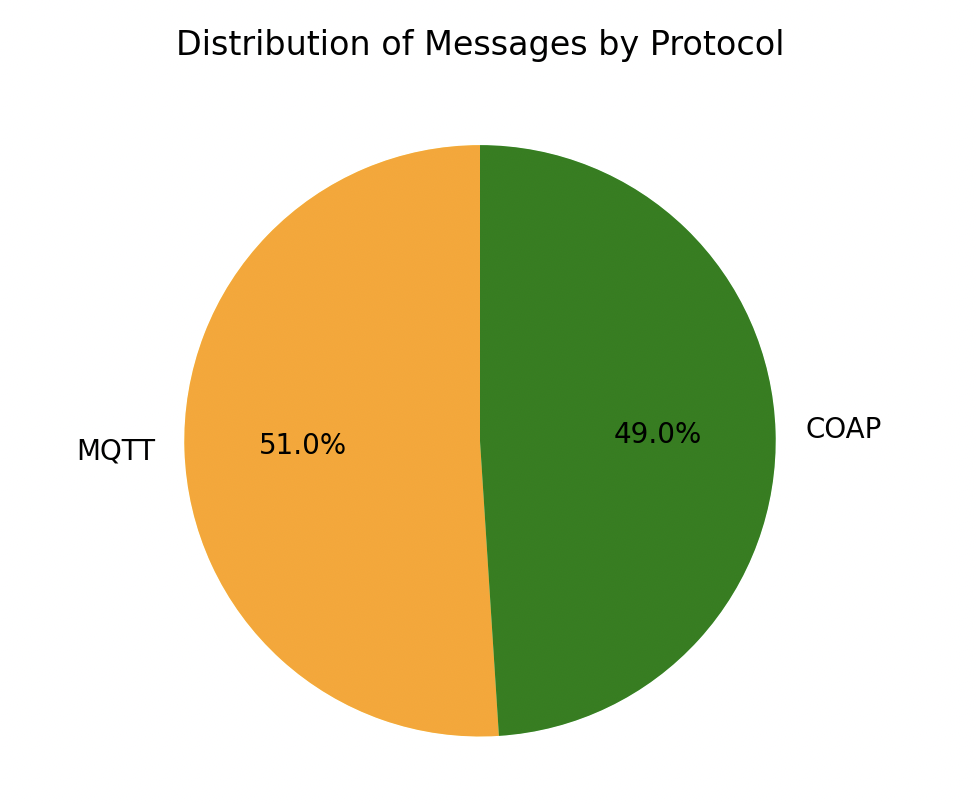}
   \caption{Distribution of Received Messages, 5 Seconds}
   \label{fig:distribution5s}
\end{figure}
\begin{figure}[ht] 
   \centering
   \includegraphics[width=0.5\textwidth]{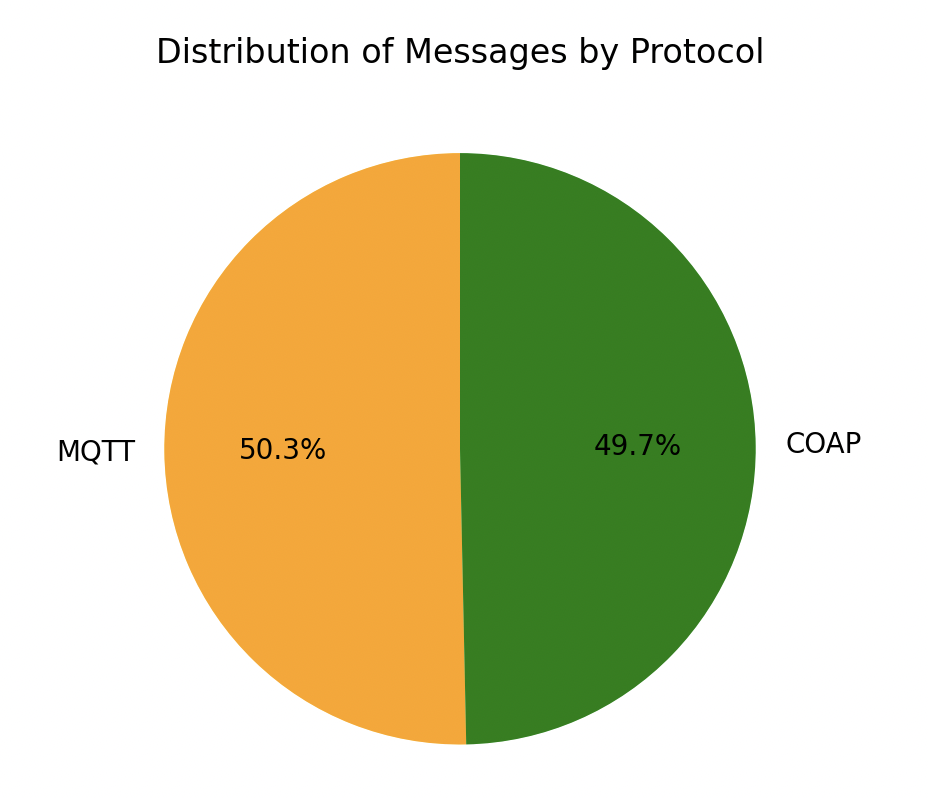}
   \caption{Distribution of Received Messages, 30 Seconds}
   \label{fig:distribution30s}
\end{figure}
Figures \ref{fig:distribution1s} through \ref{fig:distribution30s} show the distribution of messages received by the database for both communication protocols over various periods. It can be seen in \ref{fig:distribution1s} that after the first round of messages transmitted by MQTT and CoAP devices, the distribution is split 50/50, meaning the database received the same number of MQTT messages as CoAP messages. Initially, we felt that this would be the expected behavior and that the distribution of the traffic through the middleware should be 50/50. However, it can be seen in the 5-second and 30-second time spread where the number of MQTT messages received by the database was slightly higher than the number of CoAP messages. In addition, the results contradict our initial assumption that despite a difference in distribution, CoAP should have a higher throughput because it is a more lightweight protocol than MQTT. Thus, these results are significant, and more research is needed to understand these differences more thoroughly.

\section{Conclusion}
Our primary goal in researching the latency and throughput of IoT networks utilizing a middleware device was to create a platform where IoT devices that communicate differently can go through and be analyzed through a common source (middleware and database) and allow communication between heterogeneous IoT devices. Through our research, we successfully created a middleware platform that supports heterogeneous IoT communication protocols. Moreover, the fact that we could make the figures based on data from heterogeneous devices implies that data analysis through a middleware API is possible. As for communication, through our framework and gateways that access the middleware, a CoAP client can easily use the GET operator to receive information from the database inserted by an MQTT publisher and vice versa, effectively allowing communication between the heterogeneous IoT devices.

Multiple areas for future work can extend this experiment. For example, future researchers can test different variations of network distributions by sending a certain amount of messages more from the MQTT or CoAP devices. In addition, researchers can also test how the various quality of service algorithms that MQTT and CoAP have can affect latency and distribution. For example, for our MQTT simulator, we used a QoS of ``0'' meaning that the messages generated operate on a ``fire and forget'' system. This can be changed to a QoS of ``1'' signifying ``at least once'' meaning that the delivery of the message is confirmed at least one time, and finally ``2'' representing ``exactly once'' meaning there is a complex handshake that occurs that guarantees the delivery of the message only one time. Each QoS algorithm provides a different overhead, and it could be interesting to see how the latency and distribution through the middleware are affected by mixing and matching various communication protocols and their respective QoS algorithms. 


\vspace{12pt}
\end{document}